\documentclass[letterpaper, 10 pt, conference]{ieeeconf}  % Comment this line out
\IEEEoverridecommandlockouts                              % This command is only
\usepackage[usenames,dvipsnames]{xcolor}

\usepackage{graphics} % for pdf, bitmapped graphics files
\usepackage{graphicx}
\title{An Approach to Data Prefetching Using 2-Dimensional Selection Criteria}

\author{Jean Sung$^{1}$ \and Sebastian Krupa$^{1}$ \and Andrew Fishberg$^{1}$ \and Josef Spjut$^{1,2}$% <-this % stops a space
\thanks{$^{1}$Harvey Mudd College, Claremont, CA}%
\thanks{$^{2}$NVIDIA Corp., Santa Clara, CA}%
}

\begin{document}

\maketitle
\thispagestyle{empty}
\pagestyle{empty}

%%%%%%%%%%%%%%%%%%%%%%%%%%%%%%%%%%%%%%%%%%%%%%%%%%%%%%%%%%%%%%%%%%%%%%%%%%%%%%%%
\begin{abstract}

We propose an approach to data memory prefetching which augments the standard prefetch buffer with selection criteria based on performance and usage pattern of a given instruction. 
This approach is built on top of a pattern matching based prefetcher, specifically one which can choose between a stream, a stride, or a stream followed by a stride.
We track the most recently called instructions to make a decision on the quantity of data to prefetch next. 
The decision is based on the frequency with which these instructions are called and the hit/miss rate of the prefetcher. 
In our approach, we separate the amount of data to prefetch into three categories: a high degree, a standard degree and a low degree. 
We ran tests on different values for the high prefetch degree, standard prefetch degree and low prefetch degree to determine that the most optimal combination was 1, 4, 8 lines respectively. 
The 2 dimensional selection criteria improved the performance of the prefetcher by up to 9.5\% over the first data prefetching championship winner.
Unfortunately performance also fell by as much as 14\%, but remained similar on average across all of the benchmarks we tested.

\end{abstract}

%%%%%%%%%%%%%%%%%%%%%%%%%%%%%%%%%%%%%%%%%%%%%%%%%%%%%%%%%%%%%%%%%%%%%%%%%%%%%%%%
\section{Introduction}

Prefetchers are a valuable asset to any modern performance motivated processer where the performance of the memory sub-system has a large impact on overall system performance~\cite{jouppi1990improving}. 
Main memory accesses take hundreds of cycles while any on-chip memory access can be completed in a handful of cycles. Prefetching is powerful counter-measure that rarely hurts performance when implemented well~\cite{cepeda2009}. 
Specifically, prefetching occurs when a device brings data or instructions from slower memory into the cache before they are explicitly requested or needed. 
To enable efficient preemptive fetching, a strong prefetching algorithm (i.e. an algorithm that frequently predicts what may be needed in advance) must be developed and utilized. 
For example, when a memory location is requested, a trivial prefetcher may also proactively grab the data that resides in the next memory address(es)~\cite{smith1982cache}.
Ideally, this would assist in tasks that require consecutive iterative memory access.
However, for applications with other kinds of memory access patterns, this trivial prefetcher may reduce performance or cause unnecessary additional memory traffic and power consumption. 
A well performing prefetcher will only prefetch data that has a high probability of use.

\begin{figure}
  \centering
  \includegraphics[width=0.6\columnwidth]{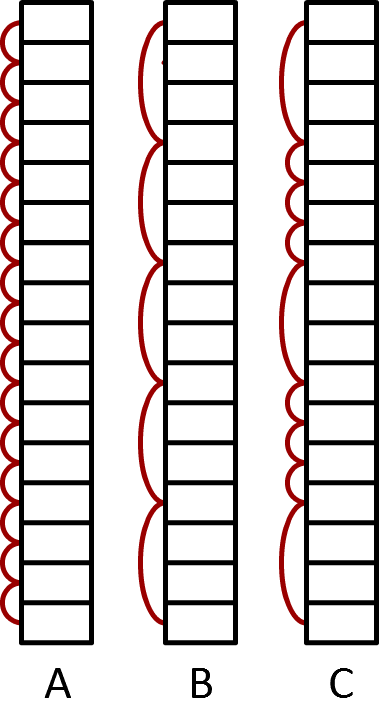}
  \caption{Memory access patterns. (A) Stream (B) Stride (C) Stream + Stride}
  \label{fig:memorypattern}
\end{figure}

Figure~\ref{fig:memorypattern} shows a few example memory access patterns that we believe to be interesting and we use in the design of our prefetcher.
Figure~\ref{fig:memorypattern} (A) shows a standard streaming memory access pattern.
A strided memory access pattern can be seen in Figure~\ref{fig:memorypattern} (B) while Figure~\ref{fig:memorypattern} (C) shows an access pattern that combines the two.
We expect that most applications will exhibit behavior that aligns with one of these patterns.
Other applications do exist which access memory in a random pattern or according to strides or other patterns that we will fail to detect when only considering these three options.

%%%%%%%%%%%%%%%%%%%%%%%%%%%%%%%%%%%%%%%%%%%%%%%%%%%%%%%%%%%%%%%%%%%%%%%%%%%%%%%%
\section{Pattern Prefetching}

\begin{figure}
  \centering
  \includegraphics[width=\columnwidth]{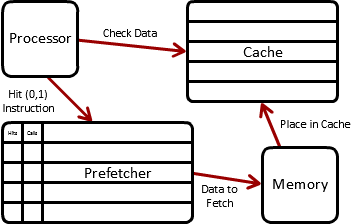}
  \caption{Prefetcher Architecture}
  \label{fig:architecture}
\end{figure}

Our prefetcher is a simple stream and stride where the degree with which it prefetches instructions is influenced by two specific metrics. These metrics are stored on a per instruction basis where we store the 128 most recently called instructions. 
%\jbs{Does the above mean our prefetcher prefetches streams and strides?}
%\jbs{What does "degree" mean here? Is that the number of lines to prefetch?}
The architecture of our prefetcher can be seen in Figure~\ref{fig:architecture}.
%64+58+26+5+32+32+6
Each entry in our instruction table consists of a 64 bit instruction pointer, the last accessed data memory address (58 bits), the stride between the last strided access (6 bits), a 5 bit stream count, 32 bits for each to count misses and cycles, and a 6 bit lru value for replacing table entries totaling 203 bits per entry. 
Since we allow for a 128 entry table, a total of 21924 bytes of storage are required for our prefetcher.
Please note that we introduce a performance optimization later in this section that increases the storage overhead slightly.

\subsection{Usage Metric}
The first metric we track is how commonly the instruction is called relative to the other instructions in the table. Each instruction is either flagged as "common" or "uncommon". We first sort all of the instructions by the number of times they have been referenced since being added to the tracker. The 78 instructions with the lowest reference count are marked as "uncommon" since they have been called infrequently relative to the other instructions.

The 50 most common instructions are considered common and the other 78 are considered uncommon.
Newly tracked instructions should initially appear as uncommon until the prefetcher sees enough references to gain confidence, which is why we have fewer instructions marked as common. 
% I rewrote the following, not sure I did any better
%We want the common list to be shorter than the uncommon list because the uncommon list will be flooded by new instructions being added to the tracker. This prevents new instructions falsely making uncommon instructions seem common. 
An alternate approach would be to tag new instructions as such before considering them in the common/uncommon sorting, but this would use space and add complexity. Furthermore, the current approach only makes it so that new instructions prefetch fewer lines of data until it is discovered if they are common or not.
Since the prefetcher wants to flood the cache, a new tag would have a similar result.

\subsection{Miss Metric}
The other metric we use is how often the instruction missed the L2 cache. This allowed the prefetcher to know the miss rates for the instructions in our buffer. We can now judge how the prefetcher influences system performance for each instruction relative to the other instructions in the buffer. If an instruction performs poorly, prefetching more data may improve performance, whereas if it performs well standard prefetching should suffice.

To decide how to quantify which instructions are doing well and which are doing poorly, we use a threshold parameter. If the threshold is 50\%, for an instruction to be considered to be doing well, it has to have a miss rate lower than 64 instructions currently being tracked. However, this is problematic since new instructions will pad the data with extreme values, i.e. an instruction called once that hit would be considered to be doing really well and one that missed would be considered to be doing really poorly. 

Instead we compare the miss rate of the current instruction to the miss rates of the 50 most common instructions along with 20 random less common instructions. We don't want to just use the most common instructions since those can have abnormally high or low miss rates, so we seed in less common instructions to alleviate this problem. This 70 entry table requires 64 bits per entry, which combines with the first required table for a total of 32768 ($2^{15}$) bits of storage.

\subsection{Usage and Miss-based Prefetching}
\begin{table}
  \centering
  \begin{tabular}{r|cc}
  usage & low miss rate & high miss rate \\ \hline
  common & standard & high \\ 
  uncommon & low & standard \\ 
  \end{tabular}
  \caption{Prefetching degree}
  \label{fig:matrix}
\end{table}

Combining the 2 metrics we get a matrix as seen in Figure~\ref{fig:matrix}. We chose to have 3 discrete prefetching degrees, a standard degree which most instructions will use, a high degree which only instructions that are common but performing poorly will use, and a low degree which uncommon instructions that are performing well will use. This should prioritize prefetching for instructions that are affecting performance most.

\begin{figure}[t!]
  \includegraphics[width=\columnwidth]{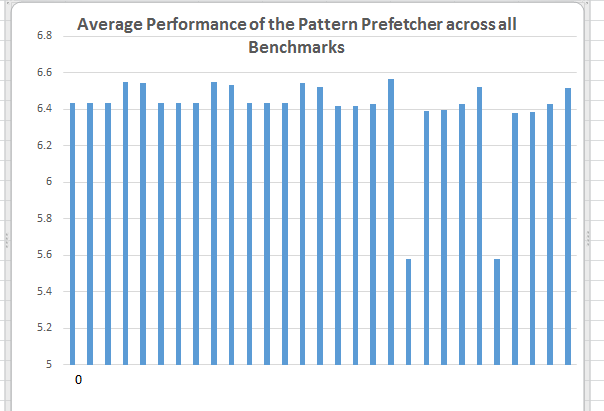}
  \caption{Graph of varying parameters used to determine optimal values for the percentage threshold as well as the three prefetching degrees. The x axis represents the variation of these parameters and the green bar represents the optimal combination of parameters. }
  \label{fig:sweep}
\end{figure}

We ran a set of tests using the simulation infrastructure described in Section~\ref{sec:results} to determine the optimal values for the three prefetching degrees as well as the threshold percentage. We tested using the provided benchmark traces as well as a set of additional benchmark traces. The optimal miss rate threshold was around 25-40 \% for all combinations of prefetching degrees. The optimal number as seen in Figure~\ref{fig:sweep} of lines to prefetch were found to be 1 for the low degree, 4 for the standard degree, and 8 for the high degree.

We found that the low degree should be as low as possible always, but never zero. This means that the benefits from a single stream or a single stride almost always outweigh the costs of prefetching an extra line of data. However, since in these case, these were instructions that were performing well, no further prefetching was necessary, thus increasing this value would only flood the cache.

The standard and high degrees both benefited from prefetching a large number of cache blocks. Increasing the standard degree up to a certain point had a slight positive increase in performance whereas increasing the high degree had a considerable increase in performance. This corroborates our hypothesis since we suspected that the common instructions that are doing poorly needed the largest amount of prefetched data.

%%%%%%%%%%%%%%%%%%%%%%%%%%%%%%%%%%%%%%%%%%%%%%%%%%%%%%%%%%%%%%%%%%%%%%%%%%%%%%%%
\section{Results}
\label{sec:results}
In order to test our prefetcher implementation, we build a prefetcher using the simulator developed by Pugsley~\cite{pugsley2015dpc2} for the Second Data Prefetching Championship.
The simulator is a trace-based simulator that models a simple out-of-order core with a 256-entry instruction window.
The simulated processor has 6-wide superscalar issue and assumes perfect branch prediction. 
The cache hierarchy configuration is found in Table~\ref{tab:caches}.
Instruction caches are not modeled and the L3 Data cache size varies with simulation. 
Data caches are simulated as inclusive and the rest of the simulation parameters can be found on the data prefetching championship website~\cite{pugsley2015dpc2}.
Note that the version of the simulator used to generate all of our results is the March 18th version that has a known bug that prevents prefetches from being promoted to demands at the memory controller.

\begin{table}
	\begin{center}
	\begin{tabular}{l|lll}
	Level & Size & Associativity & Latency\\ \hline
    L1 Data & 16KB & 8-way & 4 cycles\\
    L2 Data & 128KB & 8-way & 10 cycles\\
    L3 Data & Varies & 16-way & 20 cycles\\
    Instruction & N/A & N/A & N/A
	\end{tabular}
    \end{center}
\caption{Simulation Cache Configuration}
\label{tab:caches}
\end{table}

In order to compare our prefetcher to other styles of prefetching, we used the set of prefetchers included with the second data prefetching championship source code~\cite{pugsley2015dpc2}.
The First comparison point is called \textit{skeleton}, which represents the absence of a prefetcher, or a prefetcher that does nothing.
The second prefetcher, called \textit{next line prefetcher} simply fetches the next line whenever an address is requested.
Neither of these prefetchers requires any memory table or storage.

The \textit{stream prefetcher} works off of spatial locality by establishing a "stream" in which the program first checks if the accessed page is one of the 64 pages being kept track of. If it isn't, the oldest page is overwritten with this new page. It then checks to see if the page offset has increased or decreased. If it has increased, it checks to see which direction was already stored. If the direction of the page offset and the direction stored are the same, the confidence increases, otherwise the confidence is reset to 0 and the direction stored is flipped. It uses 1560 bytes of storage to keep track of these pages.
If the confidence is high enough (2 or greater), then the next two lines are prefetched into either the L2 cache or the LLC depending on occupancy.

The instruction pointer based stride prefetcher, called \textit{ip stride prefetcher} in the tables, keeps track of which addresses a specific instruction accessed most recently. It then assumes that future addresses to be accessed will be some fixed stride length away. The inner workings are similar to the stream, but this prefetcher stores and checks instruction pointers rather than pages. If it has seen a specific stride at between the last 3 memory addresses accessed, it prefetches the next 3 addresses that are that stride away.
It requires 32780 bytes of storage for the information it tracks.

The final point of comparison is Access Map Pattern Matching, called \textit{ampm lite} in the tables, is a simplified implementation of the winning data prefetcher from the first data prefetching championship~\cite{ishii2011access}.
We believe this is the most competitive point of comparison and will represent a high water mark for our comparison. 

\begin{figure}[t!]
  \includegraphics[width=\columnwidth]{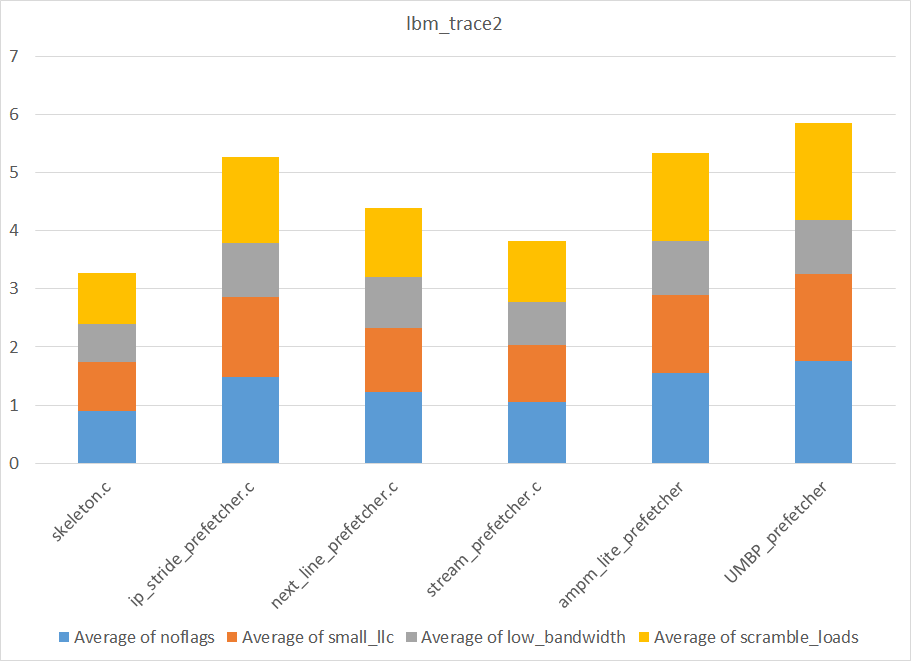}
  \caption{Best Case Scenario for Usage and Miss based Prefetcher}
  \label{fig:best}
\end{figure}

On average, the performance of the Usage and Miss-based prefetcher described in this paper was comparable to AMPM-lite. While we cannot include detailed results for every benchmark and configuration in this paper, more of the results can be found on github~\cite{sung2015github}.

\begin{figure}[t!]
  \includegraphics[width=\columnwidth]{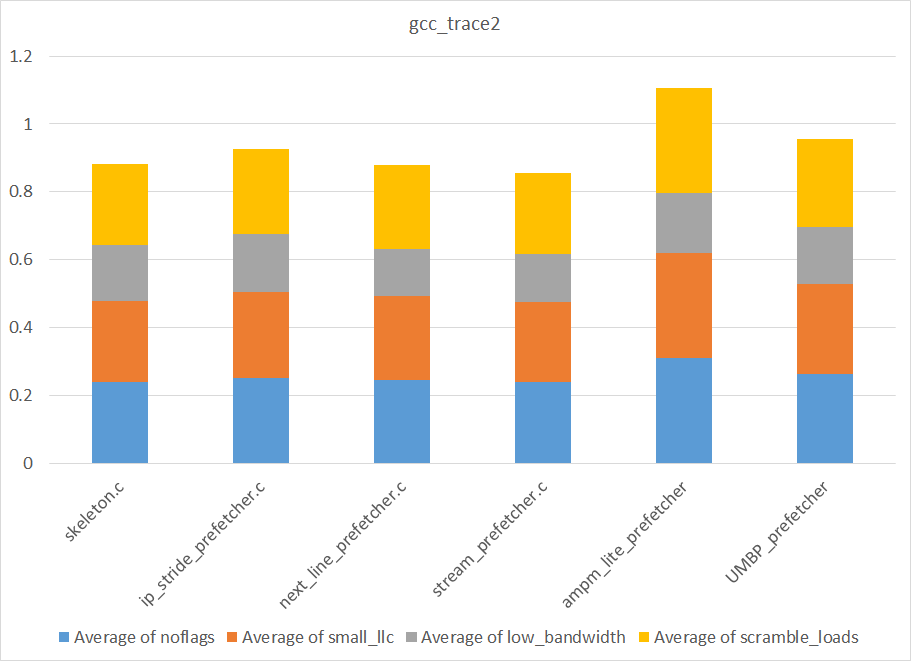}
  \caption{Worse Case Scenario for Usage and Miss based Prefetcher}
  \label{fig:worst}
\end{figure}

The prefetcher performed the best on the lbm trace, where it achieved an improvement of 9.5\% over the AMPM lite prefetcher \ref{fig:best}. The prefetcher performed poorly on gcc, where it was 14\% lower than AMPM lite \ref{fig:worst}. 
In this case, it still performed better than the other provided prefetchers. 
On most other traces, the differences were small. 
Furthermore, the better performing traces such as GemsFDTD had negligible differences between our prefetcher and AMPM lite. The worse performing prefetchers tended to have more variability.
These results imply that our prefetcher manages to handle some extreme cases which degrade performance not covered by the other prefetchers.
Likewise, AMPM manages to catch some such cases that our prefetcher misses.

\newpage
While we only present a selection of benchmark results in this paper, a larger set of simulations were run and the results can be found in the github repository~\cite{sung2015github}.
We believe that making our source code available will allow anyone a desire to reproduce our results to do so. 
Furthermore, it should be straightforward to add functionality and produce new designs from our prefetcher.

% Outline for what we got as a result :
% interesting to note: 
% for traces: lbm_trace2.dpc, leslie3d_trace2.dpc, libquantum_trace2.dpc, milc_trace2.dpc,omnetpp_trace2.dpc
% 2,4,8 with 50, 30 AND 2, 5, 8 with 50, 30
% significantly reduced performance 

% GemsFDTD_trace2.dpc
% 	our pattern prefetcher behaves uniformly across all configurations
% 	hovers around 14
% lbm_trace2.dpc
% 	less uniform, overs around 5.5 
% leslie3d_trace2.dpc
% 	pretty consistent, hovers around 3.5 
% libquantum_trace2.dpc
% 	hovers around 13, 	
% mcf_trace2.dpc
% 	hovers around 1. 2
% milc_trace2.dpc
% 	hovers around 4 
% omnetpp_trace2.dpc
% 	hovers around 7.5, 

%%%%%%%%%%%%%%%%%%%%%%%%%%%%%%%%%%%%%%%%%%%%%%%%%%%%%%%%%%%%%%%%%%%%%%%%%%%%%%%%
\section{Related Work}

The huge CPU performance gains obtained through predictive prefetching has made it an active and worthwhile research field for several decades.
Over the years, many increasingly sophisticated prefetching models have been suggested (i.e. sequential prefetcher, tagged sequential prefetcher, etc.) The most trivial prefetcher, the sequential prefetcher outlined within the Introduction, simply fetches the next line after accessing the cache~\cite{smith1982cache}.

The top three prefetchers submitted to the First Data Prefetching Championship (DPC-1)~\cite{grannaes2010}, titled AMPM (Access Map Pattern Matching), GHB-LDB (Global History Buffer - Local Delta Buffer), and PDFCM (Prefetching base on a Differential Finite Context Machine), utilized significantly more sophisticated and novel approaches to the prefetching problem.

\begin{itemize}

  \item The AMPM prefetcher~\cite{ishii2011access} divided memory into hot zones that were tracked using a 2-bit vectors to
  locate the stride zone. Our prefetcher instead tracks the last stride with a 6-bit integer. 

  \item The GHB-LDB prefetcher~\cite{dimitrov2011combining} improved upon existing prefetcher models by
  including global correlation and common stride prefetching. Our prefetcher in contrast simply tracks variable strides and streams. 
  
  \item The PDFCM prefetcher~\cite{ramos2011multi} utilized hash-based History Tables and Delta Tables to prefetch arbitrary
  degrees and distances. Our prefetcher utilizes a simpler direct look-up solution that makes prefetching decisions based
  on the frequency of instruction usage and misses.

\end{itemize}

%\jbs{qualitative comparison}

%\jbs{Cite original prefetching stuff~\cite{jouppi1990improving}}

%\jbs{Other relevant things~\cite{grannaes2010}? Maybe from list of papers that cite Jouppi.}

%\jbs{Cite AMPM~\cite{ishii2011access} and expand on background explanation. This is the DPC-1 winner.}

%\jbs{DPC1 2nd place~\cite{dimitrov2011combining}, 3rd place~\cite{ramos2011multi}.}

%\jbs{Global History Buffer~\cite{nesbit2005data}}

%%%%%%%%%%%%%%%%%%%%%%%%%%%%%%%%%%%%%%%%%%%%%%%%%%%%%%%%%%%%%%%%%%%%%%%%%%%%%%%%
\section{Conclusion}

Here we have presented a simple yet effective prefetching solution that utilizes the frequency of instruction usage and the miss rate to inform prefetching decisions. Specifically, we tune the amount of data we prefetch based on these criteria by having three prefetching degrees. Most instructions prefetch a standard amount of data, the least common instructions that have low miss rates prefetch fewer lines of data, and finally, instructions that miss a lot and are common prefetch more lines of data. 

Our prefetching solution was tested using benchmarks provided by the Second Data Prefetching Championship (DPC-2). We found that the UMBP prefetcher on average performed similarly to the provided AMPM prefetcher; however, it did perform better on traces that had lower overall IPC values. Other benchmarks showed flat performance or even some loss of performance. We believe that the approach presented here could result in more performance if more effort is spent on tuning the prefetcher.

%%%%%%%%%%%%%%%%%%%%%%%%%%%%%%%%%%%%%%%%%%%%%%%%%%%%%%%%%%%%%%%%%%%%%%%%%%%%%%%%
%\section*{APPENDIX}


\section*{ACKNOWLEDGMENT}

This work would not have been possible without the simulator written by Seth Pugsley and the support provided on the Second Data Prefetching Championship mailing list.

%%%%%%%%%%%%%%%%%%%%%%%%%%%%%%%%%%%%%%%%%%%%%%%%%%%%%%%%%%%%%%%%%%%%%%%%%%%%%%%%

\bibliographystyle{ieeetr}
\bibliography{main}

\begin{thebibliography}{1}

\bibitem{jouppi1990improving}
N.~P. Jouppi, ``Improving direct-mapped cache performance by the addition of a
  small fully-associative cache and prefetch buffers,'' in {\em Computer
  Architecture, 1990. Proceedings., 17th Annual International Symposium on},
  pp.~364--373, IEEE, 1990.

\bibitem{cepeda2009}
S.~Cepeda, ``What you need to know about prefetching,'' 2009.
\newblock
  https://software.intel.com/en-us/blogs/2009/08/24/what-you-need-to-know-about-prefetching.

\bibitem{smith1982cache}
A.~J. Smith, ``Cache memories,'' {\em ACM Computing Surveys (CSUR)}, vol.~14,
  no.~3, pp.~473--530, 1982.

\bibitem{pugsley2015dpc2}
S.~H. Pugsley, ``The second data prefetching championship,'' 2015.
\newblock http://comparch-conf.gatech.edu/dpc2/.

\bibitem{ishii2011access}
Y.~Ishii, M.~Inaba, and K.~Hiraki, ``Access map pattern matching for high
  performance data cache prefetch,'' {\em Journal of Instruction-Level
  Parallelism}, vol.~13, pp.~1--24, 2011.

\bibitem{sung2015github}
J.~Sung, S.~Krupa, A.~Fishbery, and J.~Spjut, ``Charlab dpc2 github
  repository,'' 2015.
\newblock https://github.com/charlab/dpc2.

\bibitem{grannaes2010}
M.~Grannaes, M.~Jahre, and L.~Natvig, ``Multi-level hardware prefetching using
  low complexity delta correlating prediction tables with partial matching,''
  in {\em High Performance Embedded Architectures and Compilers} (Y.~N. Patt,
  P.~Foglia, E.~Duesterwald, P.~Faraboschi, and X.~Martorell, eds.), vol.~5952
  of {\em Lecture Notes in Computer Science}, pp.~247--261, Springer Berlin
  Heidelberg, 2010.

\bibitem{dimitrov2011combining}
M.~Dimitrov and H.~Zhou, ``Combining local and global history for high
  performance data prefetching,'' {\em Journal of Instruction-Level
  Parallelism}, vol.~13, pp.~1--14, 2011.

\bibitem{ramos2011multi}
L.~M. Ramos, J.~L. Briz, P.~E. Ib{\'a}{\~n}ez, and V.~Vi{\~n}als, ``Multi-level
  adaptive prefetching based on performance gradient tracking,'' {\em Journal
  of Instruction-Level Parallelism}, vol.~13, pp.~1--14, 2011.

\end{thebibliography}

\end{document}